\documentclass[twocolumn,usenatbib]{mnras2}
\pdfoutput=1 
\usepackage{amsmath,amstext}
\usepackage[T1]{fontenc}
\usepackage{amssymb}
\input{epsf}
\usepackage{graphicx}
\usepackage{ae,aecompl}

\usepackage{hyperref}
\usepackage{amsmath}
\usepackage{amssymb}
\usepackage{mathtools}
\usepackage{bm}
\usepackage{cleveref}
\usepackage{tensor}
\usepackage{braket}
\usepackage{mhchem}
\usepackage{physics}

\usepackage{color}

\newcommand{\vect}[1]{\mathbf{{#1}}}

\newcommand{\spc}{\quad \quad \quad}

\def\be{\begin{equation}}
\def\ee{\end{equation}}
\def\beq{\begin{eqnarray}}
\def\eeq{\end{eqnarray}}

\title[Zeroth law of thermodynamics]{The zeroth law of thermodynamics in special relativity}
\author[L.~Gavassino]{L.~Gavassino\\
Nicolaus Copernicus Astronomical Center, Polish Academy of Sciences, ul. Bartycka 18, 00-716 Warsaw, Poland}
\begin{document}
\maketitle

\begin{abstract}
We critically revisit the definition of thermal equilibrium, in its operational formulation, provided by standard thermodynamics. We show that it refers to experimental conditions which break the covariance of the theory at a fundamental level and that, therefore, it cannot be applied to the case of moving bodies. We propose an extension of this definition which is manifestly covariant and can be applied to the study of isolated systems in special relativity. The zeroth law of thermodynamics is, then, proven to establish an equivalence relation among bodies which have not only the same temperature, but also the same center of mass four-velocity. 
\end{abstract}


\section{Introduction}

The modern covariant formulation of the second law of thermodynamics relies on the assumption that it is possible to define an entropy four-current $s^\nu$ with non-negative divergence, $\nabla_\nu s^\nu \geq 0$ \citep{Noto_full,Israel2009_book}. The validity of this assumption is strongly supported by relativistic kinetic theory, both classical \citep{cercignani_book} and quantum \citep{degroot_book,Florkowski:2010zz}, and by the hydrodynamics of locally isotropic fluids \citep{Gavassino2020Bulk}, including perfect fluids \citep{rezzolla_book} and chemically reacting fluids \citep{noto_rel}. This approach finds application in every branch of the relativistic hydrodynamics, including the problem of relativistic dissipation \citep{Israel_Stewart_1979,lopez2011} and multifluid hydrodynamics \citep{carter1991,CarterKhal_equivalence,andersson2007review}.

In the absence of dissipation ($\nabla_\nu s^\nu =0$), it is possible to define the total entropy of a system as the flux of the entropy current through an arbitrary spacelike hypersurface $\Sigma$ crossing the system,
\begin{equation}
S =- \int_{\Sigma} s^\nu d\Sigma_\nu.
\end{equation}
Since the result of the integral does not depend on the choice of $\Sigma$, the entropy is a Lorentz scalar, in agreement with \cite{Planck_1908} and with the microscopic statistical interpretations of the entropy \citep{Jaynes1,Jaynes2,Israel_1981_review,VanWeert1982,VonNeumannEntropy2001}. This enables us to define, starting from a theory which is necessarily local (to ensure causality), the equilibrium thermodynamics of an isolated macroscopic body, which allows to make a contact with statistical mechanics \citep{huang_book}.

The long-lasting debate on the definition of the temperature of moving bodies, traditionally called Planck-Ott imbroglio \citep{Mares2017}, originates at this point. At first \citep{Planck_1908,Tolman1933,Ott1963,Landsberg1967_cool}, the discussion was oriented in the direction of defining a transformation law of the temperature under Lorentz boosts and work in this direction is still ongoing (see \citealt{Pinto2017} for a recent review). Supporters of both Planck's and Ott's views agree on the fact that the transformation should involve a Lorentz factor, however the opinions diverge on its position (at the denominator according to \citealt{Planck_1908}, at the numerator according to \citealt{Ott1963}). Nowadays this approach has become a quest for the most natural-looking choice of variables in the differential of the energy of a moving body (see e.g. \citealt{Becattini2016,PARVAN2019}).

Other authors have approached the problem studying the equilibrium state of two weakly interacting bodies in motion with respect to each other, trying to understand whether a body ``\textit{looks hotter or colder}'' from the point of view of the other. \cite{vanKampen1968} and \cite{Israel_1981_review} have argued that in a covariant framework one must consider that the two bodies can exchange both energy and momentum and therefore the result will depend on the exact circumstance of the experiment. Following the same line of thoughts, \cite{Biro2010} have shown that, depending on the phenomenological model of heat transfer which is invoked, one can recover Planck's, Ott's or Landsberg's transformation law. \cite{Landsberg_1996,Landsberg2004} arrive at similar conclusions considering that a moving photon detector in a heat bath of black body radiation might in general measure different temperatures depending on how it averages the energies of the incoming photons with respect to the direction. Finally, \cite{Sewell2008} has studied the problem in the context of quantum statistical mechanics, proving that a body cannot satisfy the KMS conditions for different inertial frames in motion with respect to one another, arriving at the conclusion that a body has a well defined temperature only in its rest frame. 

In this paper we propose a new view on the Planck-Ott imbroglio. Using simple arguments of thermodynamics and statistical mechanics we show that the zeroth law of thermodynamics plays a fundamental role in the experimental definition of the temperature and that the deep origin of the controversy is in the fact that the operational notion of thermal equilibrium provided by standard thermodynamics (see e.g. \citealt{Guggenheim1986}) is not covariant. The purpose of our work is to revisit its definition, in a form which is manifestly covariant. Once this new formulation is provided, it will become clear that, in the absence of spontaneously broken symmetries, two bodies are in thermal equilibrium if and only if they are at rest with respect to each other and they have the same rest-frame temperature. Any condition in which two interacting bodies maintain a relative motion with respect to each other for infinite time will then be shown to be a non-ergodic system and should, therefore, be considered a metastable state, for which a complete description in the framework of equilibrium thermodynamics is not guaranteed to exist.

Throughout the paper natural units are employed: $c=\hbar=k_B=1$. For the flat space-time metric we adopt the signature $(-,+,+,+)$ and we set $\varepsilon_{0123}=+1$.

%
%

\section{Need for a zeroth law}\label{neeeeeed}

In this first section we prove that, in order to provide a unique definition for the temperature of a system exhibiting further constants of motion apart from the energy, it is necessary to specify how a thermometer is supposed to interact with the system. Our aim is to show that the concept of thermal equilibrium appearing in the zeroth law of thermodynamics relies on strict assumptions about this interaction. 


\subsection{Ambiguity of the notion of temperature}\label{ambiguo}

Under the condition that the microscopic Hamiltonian is fixed and time-independent, the second principle of thermodynamics implies that the macrostate of an isolated thermodynamic system in equilibrium is in general fully determined once all the constants of motion (fundamental, like e.g. the energy $E$, or emergent, like e.g. the winding numbers of the superfluid order parameter phases) are assigned. So, it is natural to write the entropy as a function of these variables, namely
\begin{equation}\label{Ilproblema}
S=S(E,X_A),
\end{equation} 
where the $X_A$ are $l$ conserved quantities of the dynamics of the system and we introduced the label $A=1,...,l$. Once an equation of state of this form is given, one is naturally tempted, following the standard textbook convention, to introduce the temperature through the equation
\begin{equation}\label{laprimaT}
\dfrac{1}{T} = \dfrac{\partial S}{\partial E} \bigg|_{X_A}. 
\end{equation} 
This, however, can lead to an ambiguity. In fact, if the $X_A$ are constants of motion, then this is also true for any set of quantities
\begin{equation}
Y_B =Y_B(E,X_A),
\end{equation}
with $B=1,...,l$. The $Y_B$ would, thus, constitute an equivalently good choice of variables in the construction of the equation of state, provided that the map $(E,X_A) \longrightarrow (E,Y_B)$ is one-to-one. So if we start from the equation $S=S(E,Y_B)$, then we are naturally lead to identify the temperature as
\begin{equation}\label{lasecondaT}
\dfrac{1}{T'} = \dfrac{\partial S}{\partial E} \bigg|_{Y_B} = \dfrac{1}{T} + \sum_{A=1}^l \dfrac{\partial S}{\partial X_A} \bigg|_E \dfrac{\partial X_A}{\partial E} \bigg|_{Y_B} ,
\end{equation}
which may be a priori different from the definition of $T$. So it is clear that we need to provide a more rigorous and general way to define the temperature, which must be invariant under this change of variables in the equation of state.

\subsection{The role of the thermometer}\label{TRTT}

To identify a unique definition for the temperature we need to introduce the notion of a thermometer. In standard thermodynamics an ideal thermometer is a system which interacts very weakly with the body object of study. Formally, this means that the interaction enables an exchange of heat among the thermometer and the system, but it plays a negligible role in the definition of the thermodynamic quantities and can be neglected in their computation. Thus the total energy $E_{\text{tot}}$ and entropy $S_{\text{tot}}$ can be written as 
\begin{equation}\label{EtotStot}
E_{\text{tot}} = E+E_\tau  \spc  S_{\text{tot}}=S+S_\tau,
\end{equation}
where $E_\tau$ and $S_\tau$ are respectively the energy and entropy of the thermometer. We remark that the two conditions of \eqref{EtotStot} are presented here as \textit{defining} properties of the thermometer. Their validity is necessary in thermodynamics to avoid the risk of producing further ambiguities in the definition of the temperature associated with the need of specifying the interaction \citep{Oppenheim2003} or the correlations \citep{Oppenheim2003B}. In the presence of strong gravitational fields it might become hard to ensure the additivity of the energies, however, since we are working in special relativity, the space-time is flat and it is in principle possible to make the interaction potential between the body and the hypothetical thermometer arbitrarily small.

 We impose that the only constant of motion of the ideal thermometer is the energy, so it is described by an equation of state
\begin{equation}\label{sss}
S_\tau =S_\tau (E_\tau)
\end{equation}
and there is no ambiguity in the definition of its temperature:
\begin{equation}
\dfrac{1}{T_\tau} = \dfrac{dS_\tau}{dE_\tau}.
\end{equation}
After the thermometer is put into contact with the body, they interact exchanging energy until they reach equilibrium. In equilibrium they must have the same temperature, so we can use the final value of $T_\tau$ as a measure of $T$.\footnote{An additional property of the thermometer is that its heat capacity is infinitesimal, which implies that the heat exchange does not affect the state of the body relevantly. This requirement, however, does not need to be invoked in our study. In this sense, if we replace the thermometer with its formal opposite, the heat bath, all our analysis is left unchanged.}

Let us assume that the interaction of the body with the thermometer does not break the conservation of the quantities $X_A$. Thus $E_{\text{tot}}$ and all the $X_A$ are constant during the evolution of the total system body+thermometer towards equilibrium, which is then given by the condition of maximum entropy 
\begin{equation}
\dfrac{\partial S_{\text{tot}}}{\partial E_\tau}\bigg|_{E_{\text{tot}},X_A} =0,
\end{equation}
or, equivalently,
\begin{equation}
\dfrac{1}{T_\tau} = \dfrac{\partial S}{\partial E} \bigg|_{X_A}. 
\end{equation}  
So, under the assumption that the $X_A$ are conserved by the interaction with the thermometer, we find that the thermometer measures $T$, given in equation \eqref{laprimaT}. One can show with analogous calculations that if, on the other hand, the quantities $Y_B$ were conserved by the interaction with the thermometer, then it would measure $T'$, given by equation \eqref{lasecondaT}. So, to solve the ambiguity in the definition of the temperature, one needs to find the $X_A$ which are constants of motion not only when the system is isolated, but also when an exchange of energy with a thermometer is enabled.

Note that the $X_A$ and the $Y_B$ are all simultaneously conserved in the interaction with the thermometer only if $Y_B = Y_B (X_A)$, so in the case in which $T=T'$, see equation \eqref{lasecondaT}. Therefore if the $X_A$ are conserved and $T \neq T'$, there must be at least one variable $Y_B$ whose conservation is broken due to the exchange of energy with the thermometer. 


\subsection{An unconventional thermometer}\label{auncther}

In standard thermodynamics of continuous media, the issue we presented in the previous subsection does not seem to appear because in this case there is only one independent constant of motion, apart from energy, which is the number of particles $N$. It is natural to assume that, in the conventional experimental setting in which a thermometer is in contact with a substance, the conservation of $N$ is not broken, providing a unique definition of the temperature as\footnote{Note that the volume is not a constant of motion, but it is imposed through an external potential. Therefore it is a parameter in the Hamiltonian \citep{landau5}, which is assumed from the beginning of our discussion to be fixed and time-independent. Thus, in our discussion it does not even need to appear in the equation of state and the fact that it is kept constant in the calculation of the temperature is obvious.} 
\begin{equation}
T = \bigg( \dfrac{\partial S}{\partial E} \bigg|_N \bigg)^{-1} .
\end{equation}
However it is possible to imagine unconventional thermometers which, to measure the temperature, need to break the conservation of $N$, while keeping fixed an other function $Z=Z(E,N)$. In this subsection we make a simple hypothetical example of this kind of thermometer. Our aim is to convince the reader that the ambiguity we described in this section is not merely formal, but represents a real possibility.


Let us consider an ideal gas. If we assume the particle number $N$ to be conserved, the equation of state can be written in the form
\begin{equation}
S =S(E,N)
\end{equation}
and we can define the temperature $T$ and the chemical potential $\mu$ of the gas according to the conventional prescription
\begin{equation}
dS = \dfrac{1}{T} dE - \dfrac{\mu}{T} dN.
\end{equation}
Now let us imagine to put it into contact with a thermometer, with equation of state of the form \eqref{sss}, which can interact with the gas \textit{only} by absorbing (in an annihilation process) or emitting (in a creation process) particles with a given (positive) energy $\varepsilon$. The total system gas+thermometer now is characterised by two constants of motion: the total energy $E_{tot}= E_\tau + E$ and the quantity
\begin{equation}\label{iorestoferma}
Z := E-\varepsilon N. 
\end{equation}
Thus we have built a thermometer which breaks the conservation of $N$, replacing it with the conservation of $Z$. The condition of maximum entropy with respect to the exchange of energy between the gas and the thermometer now reads
\begin{equation}
\dfrac{\partial S_{\text{tot}}}{\partial E_\tau} \bigg|_{E_\text{tot},Z} =0,
\end{equation} 
which produces the condition
\begin{equation}\label{stranissimo}
T_\tau = \dfrac{\varepsilon}{\varepsilon - \mu} T.
\end{equation}
Therefore we have proven that a thermometer of this kind, in equilibrium, has a temperature which is different from the (standard) one of the gas. 

The reader who was skeptical about the possibility of building such a thermometer without invoking the existence of any sort of Maxwell demon which selects the particles with energy $\varepsilon$ can see appendix \ref{2LT} for a very simple hypothetical example of this kind of device. There, equation \eqref{stranissimo} is proven again using pure quantum statistical mechanical arguments.

\subsection{The zeroth law of thermodynamics}\label{0thlaw}

%


One can immediately convince oneself that the thermometer that measures the temperature $T$ and the one that measures the temperature $T'$, presented in section \ref{TRTT}, cannot be considered to both reach thermal equilibrium with the system. In fact, having different temperatures, they are not in thermal equilibrium with each other and the zeroth law of thermodynamics states that \textit{the condition of thermal equilibrium is an equivalence relation}. Luckily, in the particular case of systems contained in boxes, at rest with respect to each other, the zeroth law, as it is formulated in standard textbooks, offers a simple unambiguous way to select the constants of motion $X_A$ to keep constant in the calculation of the temperature, removing any ambiguity. 

The resolution is given by the definition of the concept of thermal equilibrium appearing in the zeroth law: ``\textit{two systems are said to be in thermal equilibrium if they are linked by a wall permeable only to heat and they do not change over time}'' \citep{Caratheodory1909}. In the original formulation, walls permeable only to heat (WPOH) are fixed walls which allow the exchange of energy, but not of particles. In the context of quantum optics or high energy physics, where the interactions are mediated by particles, one cannot require the impermeability to every kind of particle, otherwise there would be no interaction at all. However we can redefine the WPOHs as fixed walls which allow the exchange of energy, but not of any Noether charge $Q_A$ arising from internal symmetries of the microscopic theory (for a general definition of conserved charge in Quantum Field Theory see \citealt{Peskin_book}).

Therefore if we make the ergodic hypothesis \citep{khinchin_book,Parisi_book_1988}, i.e. if we assume that the only constants of motion of a macroscopic system in a box are the energy and the charges $Q_A$ (which can be typically interpreted as linear combinations of particle minus antiparticle numbers), then we can write the entropy as
\begin{equation}\label{hththth}
S=S(E,Q_A),
\end{equation}
and unambiguously define the temperature as
\begin{equation}
T = \bigg( \dfrac{\partial S}{\partial E}\bigg|_{Q_A} \bigg)^{-1}.
\end{equation}
Every thermometer which is in thermal equilibrium with the system must interact in a way to conserve the amount of $Q_A$ in the system. In fact, the charge is fundamentally conserved and cannot flow from the system to the thermometer (or vice versa), because they are separated by a WPOH. Thus every thermometer measures $T$. 

The unconventional thermometer presented in subsection \ref{auncther} can never be in thermal equilibrium with the gas. In fact if we try to link it to the gas through a WPOH and we assume the particles to carry a conserved charge (e.g. a positive baryon number), then no exchange of particles is allowed, so no interaction can occur and the two bodies remain isolated with respect to each other. 

Note that Noether charges are, by construction, extensive quantities, being the flux over 3D hypersurfaces of conserved Noether currents,
\begin{equation}
Q_A = -\int_{\Sigma} J_A^\nu d\Sigma_\nu  \spc \nabla_\nu J_A^\nu =0,
\end{equation}
thus equation \eqref{hththth} also justifies the usual assumption that the entropy should be written as a function of extensive variables \citep{Callen_book}. So the zeroth law plays a fundamental role in thermodynamics, not only because it solves the problem of ambiguity of the temperature, but also because it selects the quantities which must be used as primary variables in the fundamental relation. In fact, without a zeroth law, any constant of motion of the system would be an equally acceptable free variable to be used in the equation of the entropy.

We remark that there are thermodynamic systems, such as superfluids and more in general systems with broken symmetries, which exhibit emergent constants of motion which are not conserved charges (and might not even be extensive). In these cases, however, the new constants of motion are collective quantities which can only be altered through the action of very strong interactions, while are conserved by every weak perturbation. Therefore they do not introduce any ambiguity in the definition of temperature and thermal equilibrium (see e.g. \citealt{Termo}) and will be ignored in the present discussion, even if in general they should be included in the equation of state \citep{Callen_book}.

\section{Covariant equation of state of an ergodic body}\label{coveos}


In this section we derive from first principles the covariant equation of state of an isolated system in special relativity. 
The aim is to show that the Planck-Ott controversy is simply a particular case of the general problem underlined in subsection \ref{ambiguo}. 

In a standard thermodynamic language, the term \textit{isolated} means that the system does not exchange energy with the environment, but it may exchange momentum, through e.g. the interaction with adiabatic walls. This asymmetry between energy and momentum cannot hold in a covariant formulation of thermodynamics, because we want to require that the physics (and therefore the form of the Hamiltonian) of the system are the same in any reference frame. Therefore by the term isolated we now mean that the system does not have any interaction with any external field, which otherwise would select a preferred reference frame. This immediately rules out the possibility to have an externally imposed volume, removing at the root the annoying (and not well defined) problem of the relativistic transformation of the thermodynamic pressure. 

In this context, if a system has finite size it must be self-bounded. If, for example, it is a fluid in a box, the box should be regarded as a dynamical part of the system itself, which can be accelerated, deformed or heated (it has an entropy, c.f. with \citealt{landau5}). This implies that its shape and volume is only an equilibrium property, not a parameter in the Hamiltonian.


\subsection{Deriving the equation of state}


Since, by definition, an isolated system does not have any interaction with the environment, its total four-momentum $p^\nu$ is necessarily conserved. The norm of the four-momentum is the mass of the system,
\begin{equation}\label{Massa}
M = \sqrt{-p^\nu p_\nu},
\end{equation}
and we assume it to be different from zero.

For an isolated system also the angular momentum tensor $J^{\nu \rho}$ is conserved, so special relativity provides automatically other 6 constants of motion which a priori should be included as thermodynamics variable. However, if we go to the reference frame in which $p^\nu = M \delta\indices{^\nu _0}$, we find that \citep{Weinberg_book_1972}
\begin{equation}\label{Mmmm}
J^{\nu \rho} =
\begin{bmatrix}
  0  & -M x_{CM}^1 &  -M x_{CM}^2 &  -M x_{CM}^3  \\
   M x_{CM}^1 & 0 & W^3/M & -W^2/M  \\
   M x_{CM}^2 & -W^3/M & 0 & W^1/M \\
   M x_{CM}^3 & W^2/M & -W^1/M & 0
\end{bmatrix},
\end{equation} 
where $\textbf{x}_{CM}$ is the position of the center of mass (or, more precisely, ``\textit{center of energy}'') of the system,
while
\begin{equation}
W_\mu = -\dfrac{1}{2} \epsilon_{\mu \nu \rho \sigma} J^{\nu \rho} p^\sigma
\end{equation}
is the Pauli-Lubanski pseudovector (proportional to the \textit{spin} of the body, intended as the irreducible part of its angular momentum).
Using the invariance of the laws of physics under global translations, we can set the origin in a way that $\textbf{x}_{CM}=0$, so we have that
\begin{equation}
J^{\nu \rho} =\dfrac{1}{M^2} \epsilon\indices{^\nu ^\rho _\mu _\sigma} p^\mu W^\sigma.
\end{equation}
Thus, considering also that it must be true that
\begin{equation}
W_\nu p^\nu =0,
\end{equation}
the physically relevant (in the construction of the equation of state of a body) constants of motion which the conservation of the tensor $J^{\nu \rho}$ introduces are only three out the four components of the Pauli-Lubanski pseudovector.

We call a body \textit{ergodic}, in analogy with subsection \ref{0thlaw}, if the remaining constants of motion are just the Noether charges $Q_A$ of the underlying field theory. We assume that these charges transform as scalars under Lorentz transformations. This is the case for normal continuous media, where they can be typically interpreted as particle (minus antiparticle) numbers. Recalling that the physics of the system is invariant under Lorentz transformations and that the entropy is a scalar, the most general equation of state that an ergodic body can have has, therefore, the form
\begin{equation}\label{gringo}
S = S(\sqrt{-p^\nu p_\nu}, \sqrt{W^\nu W_\nu},Q_A).
\end{equation}
This result could be equivalently proved in the context of a quantum theory as follows: since the von Neumann entropy is invariant under unitary transformations, it is a Poincar\'{e} invariant and therefore it can be written as a function of the Poincar\'{e} invariants of the theory, which are the arguments appearing in the equation of state given above.

Equation \eqref{gringo} generalizes the formula of \cite{Noto_full}, valid for static systems, to the case of rotating objects.

\subsection{The Planck-Ott controversy reinterpreted}

Since our aim is to understand only the role of the motion on the definition of the temperature we assume for simplicity that the Pauli-Lubanski pseudovector is zero and remains such also in the interaction with the thermometer. We also require that no exchange of Noether charges $Q_A$ is allowed between the body and the thermometer, in agreement with the discussion of subsection \ref{0thlaw}. Thus all the arguments in the equation of state \eqref{gringo} apart from the first one are fixed and can be ignored, leaving
\begin{equation}\label{Quellabuona}
S=S(\sqrt{-p^\nu p_\nu}).
\end{equation}
Taking the differential of this equation of state we obtain
\begin{equation}\label{tuttoqui}
dS = -\beta_\nu dp^\nu,
\end{equation}
with
\begin{equation}\label{vnidfovkm}
\beta^\nu = \dfrac{u^\nu}{T_{RF}},
\end{equation}
where we have introduced the \textit{rest-frame temperature} through the equation
\begin{equation}\label{TRFFF}
\dfrac{1}{T_{RF}} := \dfrac{d S}{dM},
\end{equation}
and the normalised four-velocity of the center of mass
\begin{equation}
u^\nu := \dfrac{p^\nu}{M} .
\end{equation}
Now the origin of the controversy becomes clear: the system, as a result of covariance requirements, is naturally given in the form \eqref{Ilproblema}, with $l=3$. In fact, chosen a reference frame, apart from the energy $E=p^0$, there are other three constants of motion implied by the conservation of the linear momentum. If we choose to write 
\begin{equation}
S=S(E,p_j),
\end{equation}
where from now on $j=1,2,3$, then equation \eqref{tuttoqui} implies
\begin{equation}\label{zigtur}
dS = \dfrac{u^0}{T_{RF}} dE - \dfrac{u^j}{T_{RF}} dp_j
\end{equation}
and we are naturally lead to define the temperature according to Planck's prescription:
\begin{equation}\label{Planck}
T = \dfrac{T_{RF}}{u^0}.
\end{equation}
If, on the other hand, we choose to write
\begin{equation}
S=S(E,v^j),
\end{equation}
where
\begin{equation}
v^j=\dfrac{p^j}{p^0}
\end{equation}
is the three-velocity of the center of mass of the system, then equation \eqref{tuttoqui} implies
\begin{equation}
dS= \dfrac{1}{u^0 \, T_{RF}} dE - \dfrac{ E \, u_j}{T_{RF}} \, dv^j.
\end{equation}
This would lead to Ott's prescription for the temperature
\begin{equation}\label{Ott}
T'= u^0 \, T_{RF}.
\end{equation}
So the ambiguity in the definition of the temperature in covariant thermodynamics is a particular case of the general problem exposed in subsection \ref{ambiguo}.


\subsection{A third option}\label{third}

Note that $p_j$ and $v^j$ are not the only two possible examples of constants of motion we could choose. Since any triplet of independent functions $Z_j= Z_j(E,p_1,p_2,p_3)$ could be put in the equation of state, there might even be definitions in disagreement with both Planck's and Ott's prescriptions.

For example, in analogy with the unconventional thermometer proposed in subsection \ref{auncther}, we can imagine the following situation. Let us assume a photon gas to be the body of which we want to measure the temperature. Chosen a reference frame, we build a hypothetical thermometer which is kept at rest in this frame by an external force\footnote{Since the external force acts on the thermometer and not on the body, it does not break the covariance of internal Hamiltonian of the body, therefore the equation of state \eqref{Quellabuona} remains valid.} and which can exchange energy only by absorbing or emitting photons with a fixed four-momentum 
\begin{equation}
q=(\varepsilon,\varepsilon,0,0),  \spc  \varepsilon >0 . 
\end{equation}  
Then, the conserved quantities of the system gas+thermometer are $E_{tot}$ and the triplet
\begin{equation}
Z_1 =  p_1 - E \spc Z_2 = p_2  \spc  Z_3 = p_3.
\end{equation} 
With calculations which are analogous to those performed in subsection \ref{auncther}, with the aid of equation \eqref{zigtur}, we can prove that when the equilibrium is reached the thermometer reports a temperature
\begin{equation}\label{bgbgbgb}
T_\tau = \dfrac{1}{ 1-  v^1} \dfrac{T_{RF}}{u^0},
\end{equation}
meaning that it measures a Doppler-shifted temperature, which is different from both \eqref{Planck} and \eqref{Ott}.

So, we have clarified that the real source of ambiguity is not special relativity by itself, but the conservation of the total linear momentum, which introduces three new constants of motion in the equation of state.

\section{Thermometers in relativity}\label{qpwodioemfowemfmo}

In subsection \ref{0thlaw} we have shown how the standard formulation of the zeroth law of thermodynamics can be used to solve the ambiguity presented in \ref{ambiguo}. However, that formulation was referring to the case in which the different bodies are linked by a WPOH, which is clearly an external field, whose presence breaks the covariance of the theory and the conservation of the momentum itself. For these reasons we cannot rely on it in our case. Furthermore, the definition of thermometer we introduced in section \ref{TRTT} is not well suited for covariant thermodynamics because the relation \eqref{sss} is not covariant, but defines an object which must be kept at rest in a given frame by the action of an external force, as in the case presented in subsection \ref{third}.

In this section we outline the steps of an ideal measurement of temperature in a context in which the physics of the body and of the thermometer are the same in any reference frame.


\subsection{Operational definition of temperature}\label{operational}

If we assume the thermometer to be a physical object (which does not break the covariance of the theory) with the least amount of macroscopic degrees of freedom possible, we need to impose it to have an equation of state of the form \eqref{Quellabuona}:
\begin{equation}\label{brutale}
S_\tau =S_\tau (M_\tau), 
\end{equation}
with $M_\tau$ the mass of the thermometer, given by
\begin{equation}
M_\tau = \sqrt{-p_\tau^\nu p_{\tau \nu}},
\end{equation}
where $p_\tau^\nu$ is its four-momentum. Let us consider the following idealized experiment:
\begin{itemize}
\item A thermometer is prepared with an initial mass and located inside the system in a not specified state of motion (formally, this means that the four-momentum $p_\tau^\nu$ in the initial conditions is arbitrary)
\item A weak interaction between the body and the thermometer is switched on and the whole system evolves towards its equilibrium state (formally, this is obtained maximizing the total entropy of the full system body+thermometer compatibly with the constants of motion, which are assigned in the initial conditions).
\item After the state of equilibrium is reached, the interaction is switched off and the thermometer is extracted from the body. An experimenter, then, weighs the thermometer accurately, without altering its internal state, obtaining a measure of its mass and therefore, using \eqref{brutale}, of its (rest-frame) temperature: 
\begin{equation}
T_\tau = \bigg( \dfrac{dS_\tau}{dM_\tau} \bigg)^{-1}.
\end{equation} 
\end{itemize}

The experiment we presented constitutes the ideal temperature measurement procedure in special relativity. Note that the experimental setting presented in subsection \ref{third} is naturally ruled out in a fully covariant context. In fact the thermometer, absorbing photons, exchanges momentum with the gas, and therefore should accelerate. The only way to prevent it is to admit the existence of an external force which counteracts this effect, breaking the covariance of the theory.

\subsection{The constants of motion}\label{comot}

In the experimental procedure outlined in the previous subsection, the equilibrium state that the isolated system body+thermometer reaches at the end of a transient is the one which maximizes
\begin{equation}
S_{\text{tot}}= S(\sqrt{-p^\nu p_\nu}) + S_\tau (\sqrt{-p_\tau^\nu p_{\tau \nu}}).
\end{equation}
Thus, we need to impose, see equation \eqref{vnidfovkm},
\begin{equation}\label{ioelei}
\delta S_{\text{tot}} = -\beta_\nu \delta p^\nu -\beta_{\tau\nu} \delta p_\tau^\nu =0 
\end{equation}
for every perturbation of the 8 variables $p^\nu$, $p^\nu_\tau$ allowed by the conservation of the constants of motion. The conservation of the total four-momentum 
\begin{equation}
p_{\text{tot}}^\nu = p^\nu + p^\nu_{\tau},
\end{equation}
however, implies that 
\begin{equation}
\delta p^\nu = -\delta p_\tau^\nu,
\end{equation}
so equation \eqref{ioelei} becomes
\begin{equation}\label{frz}
(\beta_\nu - \beta_{\tau \nu}) \delta p^\nu_\tau =0.
\end{equation}
Thus, we immediately obtain that if the thermometer is able to exchange freely energy and momentum with the body, then the variations $\delta p^\nu_\tau$ are completely free and equation \eqref{frz} implies
\begin{equation}\label{justyou}
u^\nu_\tau = u^\nu  \spc  T_{\tau} = T_{RF} .
\end{equation}
The first equation is the thermodynamic explanation of the friction as an entropic force. If the two bodies are free to exchange both energy and momentum, the exchange will lead them to a final state in which they comove and have the same rest-frame temperature.

The only way to avoid the trivial outcome of a thermometer comoving with the body is to impose the presence of other constants of motion, which can force the existence of relative motions also in equilibrium. The easiest way to do this consists of imposing that the four-velocity $u_\tau^\nu$ of the thermometer is conserved in the interaction with the body (as has been proposed by \citealt{Becattini2016}). If this is the case, then the variation $\delta p_\tau^\nu$ in \eqref{frz} are no more completely free, but satisfy
\begin{equation}\label{PPPPPPlan}
\delta p_\tau^\nu = u_\tau^\nu \delta M_\tau .
\end{equation}
Therefore, the condition for the entropy to be maximum with respect to $M_\tau$, with $p_{\text{tot}}^\nu$ and $u_\tau^\nu$ constant, reads
\begin{equation}
\dfrac{u_\nu u_\tau^\nu}{T_{RF}} + \dfrac{1}{T_\tau} =0,
\end{equation}
which in the rest frame of the thermometer becomes Planck's prescription, equation \eqref{Planck}. However, this is only one possibility. One could instead design a situation in which the thermometer does not alter the state of motion of the body. In this case the constants of motion would be the components of $u^\nu$, giving
\begin{equation}\label{OOOOOOttt}
\delta p_\tau^\nu = -u^\nu \delta M .
\end{equation}
The final condition would read
\begin{equation}
\dfrac{1}{T_{RF}} + \dfrac{u_\nu u_\tau^\nu}{T_\tau} =0,
\end{equation}
in agreement with Ott's prescription, equation \eqref{Ott}. There are, clearly, other possible assumptions for $\delta p_\tau^\nu$ which lead to alternative equilibrium conditions.

We have shown that the result of a temperature measurement can vary from thermometer to thermometer if the we admit the presence of a relative motion in equilibrium. Everything depends on which energy-momentum exchanges are allowed by the interaction and this, in turn, depends on how the thermometer is designed. This formalizes from a statistical perspective the results of \cite{Biro2010}, who have shown that different phenomenological models of heat transfer (i.e. different directions of allowed four-momentum exchanges) lead to different equilibrium states.

It is, however, clear that a constraint of the kind \eqref{PPPPPPlan} or \eqref{OOOOOOttt} represents a non-ergodicity of the system body+thermometer which is unlikely to arise in realistic situations where a form of friction, even if small, always exists.\footnote{Again we remark that in this work systems with spontaneously broken symmetries, like superfluids, are not considered because the role of their emergent constants of motion in the definition of the temperature has already been clarified. The systems under consideration are only the ergodic bodies presented in subsection \ref{coveos}.} For the interested reader, we propose in appendix \ref{Iosonoegodico} an ideal system which seems, at first, to allow for a permanent relative motion in equilibrium. Then we show how nature finds a way to prevent its existence.

\subsection{Indirect thermometers}

Motivated by the work of \cite{Landsberg_1996}, we need to clarify a point before moving on with the general discussion. In the context of thermodynamics, a rigorously defined thermometer is an object which always reports only its own temperature. Its operating mechanism is based on the zeroth law and not on the knowledge of the microphysics of the body. To clarify the importance of this distinction we give the following example.

Let us suppose that we need to measure the temperature of a photon gas in equilibrium, at rest with respect to us. Since the number of photons is not conserved, all the local properties of the gas depend on a single parameter. This implies that the average value of any intensive observable can be written as a function of the temperature. The famous equation of state of the radiation gas
\begin{equation}\label{aT4}
\varepsilon = a \, T^4,
\end{equation}
where $\varepsilon$ is the energy density and $a$ is the radiation constant, is an example. As a result, any device which measures an observable whose average value is monotonic in the temperature can be used as an indirect thermometer. We take as an example a hypothetical instrument which measures the energy density and then is calibrated to report on a screen the quantity
\begin{equation}
T_{m} := (\varepsilon/a)^{1/4}.
\end{equation}
Clearly, if equation \eqref{aT4} holds, we have $T_m=T$, therefore this seems to be a reliable thermometer. Now let us imagine that the conditions in the environment change and suddenly all the interactions conserve the number of photons. Now the gas can have in principle a finite chemical potential $\mu$ and will have an equation of state
\begin{equation}
\varepsilon = \varepsilon(T,\mu).
\end{equation}  
If now we try to measure the temperature with the device we presented above we may get a wrong answer, because of the additional dependence of $\varepsilon$ on the chemical potential.

So we have shown that there is no guarantee that a device which is calibrated to associate a temperature to the mean value of a different observable will give the correct result when unexpected constants of motion arise. This is the explanation of the non-uniqueness of the outcome of the temperature measurement proposed by \cite{Landsberg_1996} in their thought experiment. In fact, they consider a detector which is calibrated to associate a temperature to the spectral properties of the photon gas at rest and they imagine to use it in a situation in which the total momentum is conserved and different from zero.  

In this work we are not dealing with this kind of devices, but only with the ideal thermometers we presented in subsection \ref{operational}.

\section{Covariant formulation of the zeroth law}

We are finally able to revisit the concept of thermal equilibrium, presented in subsection \ref{0thlaw}, and adapt it to a relativistic context. As a result, we will also show that it is achieved only when the bodies are at rest with respect to each other.

\subsection{A new notion of thermal equilibrium}\label{nofhermalequilibrium}

Since any kind of externally imposed wall breaks the covariance of the theory, we need to revisit the notion of thermal equilibrium replacing the WPOH with a more general interaction. From a microscopic perspective, the wall only permeable to heat is an idealization invoked to describe a condition in which the two systems are linked by a very weak interaction which does not allow the exchange of conserved charges (of the kind described in subsection \ref{0thlaw}) among the bodies. Formally, this means that the total Hamiltonian $H$ is the sum of the Hamiltonians of the individual systems, plus a small interaction potential $V_I$ such that
\begin{equation}\label{giggione}
\comm{Q_A^{(i)}}{V_I} =0 \spc \forall A,i,
\end{equation}
where $Q_A^{(i)}$ is the amount of conserved charge $A$ contained in the system $i$. To make this theory covariant and compatible at a fundamental level with an underlying Quantum Field Theory, we can require the interaction to have a form
\begin{equation}\label{interazione}
V_I = -\int \mathcal{L}_I \, d_3 x,
\end{equation} 
where $\mathcal{L}_I$ is a scalar and constitutes the interaction term of the Lagrangian density (function of the fields and not of their derivatives). In general, the term $\mathcal{L}_I$ will be a coupling of the two bodies with some particles which are exchanged by the systems and therefore mediate the interaction. It is, then, clear that the condition \eqref{giggione} is fulfilled, provided that the mediator does not carry any conserved charge. Thus we can formulate the covariant notion of thermal equilibrium in special relativity as follows: ``\textit{two systems are said to be in thermal equilibrium if they weakly interact through charge-neutral (in the sense of Noether's theorem) mediators and there is an inertial frame in which they are both in a stationary state}''.

Note that, even if this new definition has been derived in a fully covariant context, it can still be applied to those cases in which external agents (like walls) are included. In this sense it represents only a generalization of the previous one, able to include the problem of the moving objects. In fact an ideal wall only permeable to heat is formally described as a surface which acts as a perfectly reflecting wall for baryons and leptons, but which is permeable to photon radiation (a photon does not carry any charge, being its own antiparticle).

\subsection{The zeroth law in special relativity}

Now that an extended notion of thermal equilibrium has been proposed, we are able to address from a statistical perspective the question about whether or not the zeroth law can hold also in a fully covariant context.

When two bodies exchange a mediator, this will transfer its four-momentum, which, if the sender body is macroscopic, will be in general randomly oriented and its frequency will take values in a continuous spectrum. Therefore we can conclude that in equilibrium the entropy is maximized with respect to any infinitesimal exchange of four-momentum. Formally this means that we need to take $\delta p_\tau^\nu$ arbitrary in equation \eqref{frz}, leading to equation \eqref{justyou}, which we present again here for two arbitrary bodies $a$ and $b$:
\begin{equation}
u^\nu_a = u^\nu_b \spc  T_{RF}^a = T_{RF}^b .
\end{equation}
If now we go to the common rest-frame of the two bodies, they are both in a stationary state, proving that indeed they are in thermal equilibrium, according to the definition proposed in the foregoing subsection. 

It is in principle not impossible to design ad hoc interactions which make one of the bodies opaque only to one specific value of four-momentum of the mediator and transparent to all the others, producing a constraint on the exchange of four-momentum $\delta p_\tau^\nu$ in equation \eqref{frz}. However, considering that at a fundamental level the interaction is not arbitrary, but has the form \eqref{interazione}, we can assume that in realistic situations, if we take into account the processes at all the orders, this phenomenon will be in the end broken  (in agreement with the thought experiment proposed in appendix \ref{Iosonoegodico}). As we explained in subsection \ref{comot}, this assumption corresponds to a statement of non-existence of additional constants of motion apart from the Noether charges of the theory. We can, thus, consider it a manifestation of the ergodic hypothesis.

We have shown that in covariant thermodynamics two bodies are in full thermal equilibrium if they have the same rest-frame temperature and are at rest with respect to each other. This immediately implies that the condition of being in thermal equilibrium is still an equivalence relation, whose equivalence classes are all the sets of bodies with the same $\beta^\nu=T_{RF}^{-1} u^\nu$. The zeroth law of thermodynamics is, then, still valid.

\section{Applications}

Below we propose three simple applications of our study.

\subsection{Heat baths in special relativity and the covariant free-energy principle}

A heat bath is a body whose macroscopic properties are not relevantly altered by the interaction with the system object of study. In standard thermodynamics this is ensured by imposing an infinite heat capacity. It is clear that in special relativity, when both exchanges of energy and momentum are included, we need to provide a specification also of its inertia. We therefore define a heat bath as body with (effectively) infinite heat capacity and mass. Its equation of state needs to be expressed in a covariant form as a relationship between its mass $M_H$ and entropy $S_H$ (see equation \eqref{Quellabuona}), therefore we may write an expanded effective fundamental relation 
\begin{equation}\label{Bagnooo}
S_H =S_{H0} + \dfrac{M_H}{T_H} ,
\end{equation}
where $S_{H0}$ and $T_H$ are two constants. The quantity $T_H$ is the rest-frame temperature and the fact that it is a constant guarantees the divergence of the heat capacity.  

When an interaction with a system (whose thermodynamic variables will not carry any label, as usual) is enabled, system and heat bath will exchange energy and momentum, conserving the total four-momentum
\begin{equation}\label{quattromome}
p_{tot}^\nu = p_H^\nu +p^\nu.
\end{equation}
Thus, considering that 
\begin{equation}
p_H^\nu = M_H u_H^\nu,
\end{equation}
where $u_H^\nu$ is the center of mass four-velocity of the bath, we obtain that
\begin{equation}
\delta u_H^\nu = - \dfrac{\delta p^\nu +u_H^\nu \delta M_H}{M_H}.
\end{equation} 
In the limit in which $M_H \longrightarrow +\infty$, with finite exchanges of four-momentum, 
\begin{equation}
\delta u_H^\nu \longrightarrow 0,
\end{equation}
so the heat bath does not accelerate. As a result, a heat bath in special relativity is an object with constant temperature and center of mass four-velocity. The manifestly covariant formulation of this statement is
\begin{equation}
\beta_H^\nu = \dfrac{u_H^\nu}{T_H} = const.
\end{equation} 

Let us study the condition of equilibrium of a system in contact with a heat bath. The law of non-decreasing entropy reads
\begin{equation}
\delta S_{tot} = \delta S_H + \delta S \geq 0.
\end{equation}
With the aid of equation \eqref{Bagnooo} we can rewrite this condition as
\begin{equation}
\delta (T_H S_{tot}) = -u_{H\nu} \delta p_H^\nu + T_H \delta S \geq 0.
\end{equation}
Invoking the conservation of the total four-momentum we find
\begin{equation}
\delta (T_H S_{tot}) = u_{H\nu} \delta p^\nu + T_H \delta S \geq 0.
\end{equation}
Since $u_{H\nu}$ and $T_H$ are constant, they can be brought inside the variation, producing the condition
\begin{equation}\label{nonincreasing}
\delta F \leq 0,
\end{equation}
with
\begin{equation}\label{freeenergy}
F = -u_{H\nu}p^\nu -T_H S.
\end{equation}
This is the covariant formulation of the principle of minimum free energy for a system in contact with a heat bath. Note that in the reference frame of the bath the expression for the covariant free energy reduces to
\begin{equation}
F = E -T_H S.
\end{equation}
Thus we have proven that the equilibrium state of a system in contact with a heat bath is the one which minimizes the Helmholtz free energy (over the manifold of states with $T_{RF}=T_H$) measured in the rest-frame of the heat bath.

This is in agreement with the quantum statistical result of \cite{Sewell2008}, who proved that a body that serves as a heat bath (in the sense of the zeroth law) can obey the KMS conditions only in its own rest-frame, making equation \eqref{freeenergy} the only possible covariant generalization of the free energy.

Finally we note that if we search for the equilibrium state of the system imposing 
\begin{equation}
\delta F =0,
\end{equation}
we obtain the conditions
\begin{equation}
u^\nu = u_H^\nu  \spc T_{RF} =T_H,
\end{equation}
in accordance with our formulation of the zeroth law.

\subsection{Relativistic unification of heat and friction} 

The root of the Planck-Ott imbroglio was the disagreement on the relativistic transformation of the heat \citep{Pinto2017}. The two views, and a simple way of deriving them, are summarized in appendix \ref{Calormio}. Now we are able to clarify this issue.

In standard thermodynamics, the work exerted on a system is its change of energy due the time-dependence of external macroscopic forces. Following \cite{landau5}, a macroscopic force can be modelled as an external field which appears in the microscopic Hamiltonian of the system and the work it makes is the result of a time-dependence of it. In fact, if this field changes with time, this breaks the invariance of the microscopic theory under time translations and, as a result, the (Noether) conservation of the energy of the body.

If we follow this interpretation, the generalization to special relativity is straightforward. Let us consider a system in contact with a heat bath and interacting with an external field (which does not act on the bath directly). From \eqref{quattromome} we find
\begin{equation}\label{quid}
\delta p_{tot}^\nu =\delta p_H^\nu +\delta p^\nu.
\end{equation}
The presence of the field now breaks the invariance of the theory under the whole Poincar\'e group, because the field might have an arbitrary dependence both on space and time. Therefore the total four-momentum is not conserved and we call its variation work four-vector:
\begin{equation}
\delta \mathcal{W}^\nu :=\delta p_{tot}^\nu \neq 0 .
\end{equation}

Note that if we consider the whole universe (external field included) as a bigger system, and the field becomes a dynamical variable itself, then the Poincar\'e invariance is restored and the four-momentum of the universe is again conserved. Thus $-\delta \mathcal{W}^\nu$ represents the four-momentum transferred from the body to the external environment mediated by the action of the field, in agreement with the common practical interpretations.

Considering that the heat bath does not interact directly with the field, its variation of four-momentum can only be the outcome of a transfer to the body, which allows the introduction of a heat four-vector
\begin{equation}\label{QQQwerty}
\delta \mathcal{Q}^\nu := -\delta p_H^\nu.
\end{equation}
As a result, equation \eqref{quid} becomes the first law in special relativity:
\begin{equation}\label{envoecm}
\delta p^\nu = \delta \mathcal{Q}^\nu + \delta \mathcal{W}^\nu.
\end{equation}
Formally, this four-vectorial construction reminds the one proposed by Ott (see appendix \ref{Calormio}), but the separation into contributions is completely different. It is interesting to study two particular cases which will clarify its physical meaning and show its consistency with the standard definitions.

Let us consider the case in which no interaction with the heat bath occurs. From \eqref{QQQwerty} we have $\delta \mathcal{Q}^\nu=0$. Now if we choose a reference frame and decompose \eqref{envoecm} into its time and space components we obtain the equations
\begin{equation}
\delta E = \delta \mathcal{W}^0  \spc  \delta p^j= \delta \mathcal{W}^j.
\end{equation}
Let us assume that, in this reference frame, the external field does not depend on time. Then the invariance under time translations is restored and the energy of the body is conserved, giving $\delta E = \delta \mathcal{W}^0 =0$. However, if the field depends on space, the momentum is not conserved and in principle $\delta \mathcal{W}^j \neq 0$. This is the case, for example, of a gas enclosed in a box with perfectly reflecting walls. These ideal walls are nothing but an external stationary potential which break the conservation of the momentum of the gas. If, on the other hand, the walls start moving slowly, then also the energy can change, and it does according to the well known formula $\delta E=-P \delta V$, where $P$ and $V$ are respectively pressure and volume. Thus in this case we have
\begin{equation}
\delta \mathcal{W}^0 = -P \delta V.
\end{equation}
So we hope we have convinced the reader that the four-vector $\delta \mathcal{W}^\nu$ is the natural relativistic generalization of the thermodynamic work.

Now we can focus on the heat. To do this we consider an opposite situation in which no external field is applied and the body interacts only with a heat bath. Then $\delta \mathcal{W}^\nu =0$ and we have
\begin{equation}\label{zozzu}
\delta p^\nu = \delta \mathcal{Q}^\nu.
\end{equation}
To study the physical interpretation of this equation, we work in the reference frame of the heat bath and write the decomposition
\begin{equation}\label{zazzu}
\delta E = \delta \mathcal{Q}^0  \spc  \delta p^j= \delta \mathcal{Q}^j.
\end{equation}
In this frame the condition of non-increasing free energy \eqref{nonincreasing} can be easily rewritten in the form
\begin{equation}\label{zizzu}
\bigg( 1- \dfrac{T_H}{T_{RF}} u^0 \bigg) \delta E + \dfrac{T_H}{T_{RF}} u_j \delta p^j \leq 0.
\end{equation}
If we impose that the body is at rest with respect to the bath, then we have $u^j =0$ and equation \eqref{zizzu}, combined with \eqref{zazzu}, becomes
\begin{equation}
\bigg( 1- \dfrac{T_H}{T_{RF}} \bigg) \delta \mathcal{Q}^0 \leq 0.
\end{equation}
This allows us to identify $\delta \mathcal{Q}^0$ with the heat exchange and we have recovered the Clausius formulation of the second law\footnote{``\textit{There is no thermodynamic transformation whose sole effect is to extract a quantity of heat from a colder reservoir and to deliver it to a hotter reservoir}'' (see e.g. \citealt{huang_book})}, which holds strictly for bodies which are at rest with respect to each other. On the other hand, if we impose that the two bodies have the same temperature $T_{RF}=T_H$ and their relative speed is non-zero, but small compared to the speed of light (i.e. $u^0 \approx 1$), we find from the combination of \eqref{zizzu} and \eqref{zazzu} the condition
\begin{equation}\label{xoxxu}
u_j \delta \mathcal{Q}^j \leq 0.
\end{equation}
However, the only force which acts on a body which is moving with respect to the environment, in the absence of external fields, is the friction with the environment $f_{F}^j$. Thus from Newton's second law we find
\begin{equation}\label{xixxu}
 \delta p^j =f_{F}^j \delta t,  
\end{equation}
where $\delta t$ is the time interval in which the exchange of momentum occurs. Plugging \eqref{xixxu} into \eqref{xoxxu} we obtain the other familiar law
\begin{equation}
u_j f_{F}^j \leq 0,
\end{equation}
which states that the friction always acts in a way to reduce the relative speed between the bodies (in agreement with our formulation of the zeroth law).

The analysis we have made shows that we can interpret the heat four-vector $\delta \mathcal{Q}^\nu$ as the joint action of friction and heat exchanges with the bath. Special relativity has dictated the fundamental non-divisibility of these two dissipative phenomena, as different components of the same four-vector. Our analysis in subsection \ref{nofhermalequilibrium} also provides a microscopic interpretation of this unification. It shows that, since any interaction is the result of the exchange of a mediator, it always involves a transfer of both energy and momentum, making any separation of the contributions merely artificial.

In conclusion, heat transforms as a four-vector in special relativity. However, its space components do not necessarily vanish in the rest-frame of the body, thus Ott's transformation law does not apply in general.

\subsection{Temperature of rotating objects}

It is well known that rotating relativistic objects in full thermodynamic equilibrium have non-uniform temperature \cite{cercignani_book}. The Killing condition for the temperature vector field implies that the local rest-frame temperature of the volume elements has a dependence
\begin{equation}
T(R) = \dfrac{T(0)}{\sqrt{1-\omega^2 R^2}},
\end{equation}
where $\omega$ is the (uniform) angular velocity of the body and $R$ is the distance from the axis of rotation. The natural question which now arises is how we can define the temperature of a rotating object which is compatible with the zeroth law. To address this issue we need to consider equation \eqref{gringo} (ignoring the presence of charges) including a non-zero norm
\begin{equation}
W = \sqrt{W_\nu W^\nu}
\end{equation} 
of the Pauli-Lubanski pseudovector. The equation of state then reads
\begin{equation}
S=S(M,W).
\end{equation}
Our intuition tells us that we can define the inverse of the temperature as the partial derivative of $S$ with respect to $M$, but again we face the problem that we do not know which variables we should keep constant. Thus, in analogy with what we did in the previous sections, we consider an interaction with an ideal (spin-less) thermometer and we search for the maximum of the function
\begin{equation}
S_{tot} =S(M,W) + S_\tau (M_\tau). 
\end{equation}
The maximum needs to be computed compatibly with the constraints of conservation of the total four-momentum and angular momentum tensor:
\begin{equation}
\begin{split}
& p_{tot}^\nu = p^\nu + p_\tau^\nu \\
& J_{tot}^{\nu \rho} = J^{\nu \rho} + J_\tau^{\nu \rho}. \\
\end{split}
\end{equation} 
If one performs the calculation of the variations explicitly, imposing
\begin{equation}
\delta S_{tot}=0  \quad \quad \delta p^\nu =-\delta p_\tau^\nu \quad \quad \delta J^{\nu \rho} = -\delta J_\tau^{\nu \rho},  
\end{equation}
they will find, as a unique result, that the thermometer has to be at rest in the center of mass of the rotating body, measuring a temperature
\begin{equation}\label{Saddle}
T_\tau = \bigg(\dfrac{\partial S}{\partial M} \bigg|_{W/M} \bigg)^{-1}.
\end{equation}
It turns out, however, that this is not the maximum of the entropy, but only a saddle point, which results from the mathematical symmetries of the problem. The entropy does not admit a maximum, but only a supremum, which can be computed using the following argument. 

It is natural to assume that
\begin{equation}\label{dsdw}
\dfrac{\partial S}{\partial W} \bigg|_M \leq 0,
\end{equation}
because, as the spin grows at constant mass, we are subtracting energy from the ``chaotic'' motions, forcing it into the collective rotation. We consider a state in which the thermometer and the body are far away from each other, in the limit in which the center of mass of the thermometer goes to infinity. If we remove the spin-angular momentum from the body entirely, setting
\begin{equation}\label{w00}
W=0,
\end{equation}
this needs to be transferred into orbital angular momentum to ensure the overall conservation. However the orbital angular momentum of the thermometer scales as
\begin{equation}
L_\tau \sim R \, p_\tau,
\end{equation}
where $R$ is the distance of the thermometer from the chosen pole (which can be set in the center of mass of the body) and $p_\tau$ is the spatial transversal component of the momentum. Thus we have that, for $R \longrightarrow +\infty$, only a negligible value of $p_\tau$ is required to provide enough angular momentum to ensure the conservation of the total. Therefore we have proven that with an infinitesimal variation of the momentum of the thermometer we can always remove the spin angular momentum from the body and impose equation \eqref{w00} without violating the conservation of the total angular momentum.  

Considering equation \eqref{dsdw}, we see that this transformation is always favourable and leads to a growth of entropy. Thus we obtain that the supremum entropy is achieved for
\begin{equation}
W=0  \spc T_\tau = T  \spc u_\tau^\nu = u^\nu,
\end{equation} 
which is reached only in the limit in which body and thermometer are infinitely far from each other. This result remarkably shows that, even if a rotating body can be, in principle, in thermodynamic equilibrium, it is impossible to associate a temperature to its state from the zeroth law if $W \neq 0$. This, obviously, does not prevent one from defining a generalised notion of temperature through, e.g., the saddle point equation \eqref{Saddle}. However, assuming this as the temperature, all the machinery of equilibrium thermodynamics is not guaranteed to apply consistently.

%

\section{Conclusions}

We have shown that the presence of other constants of motion apart from the energy in a thermodynamic system can lead to a fundamental ambiguity in the definition of the temperature. The Planck-Ott imbroglio has been explained to arise as a direct consequence of this problem, produced by the need of requiring the conservation of the total linear momentum.  

We have explained that the zeroth law of thermodynamics plays a crucial role in selecting the appropriate experimental setting which defines a temperature measurement, removing all the ambiguities. However, we have seen that the standard notion of thermal equilibrium invoked in classical thermodynamics refers to experimental conditions which break the covariance of the theory at a fundamental level and therefore has to be revisited to be applicable to the case of moving bodies in special relativity.

We have proposed a new notion of thermal equilibrium which is fully covariant and well suited for applications in both relativistic and non-relativistic contexts (provided that the gravitational field is negligible). Under the ergodic assumption we have proven that, using this definition, the zeroth law of thermodynamics is still valid and full thermal equilibrium is achieved only when all the systems have the same rest-frame temperature and are at rest with respect to each other. 

We have, then, applied our results to three selected open problems of relativistic thermodynamics. We have provided a covariant formulation of the notion of a heat bath and of the principle of minimum free energy. The result is in complete agreement with quantum statistical mechanical calculations \citep{Sewell2008}, establishing a direct connection of our approach with microphysics. Then, we have proposed a solution to the long-lasting debate about the relativistic notions of thermodynamic work and heat. Our minimal approach has been to generalize the definition provided in standard textbooks \citep{landau5} to a relativistic spacetime. This extension was constructed in a way to be rigorously defined at every scale, this making it well suited for both theoretical modeling and practical application. In the end, we studied the problem of defining the temperature of rotating relativistic objects. We found that it is not possible to attribute to them a temperature from the zeroth law. This is in accordance with the fact that, since their temperature is not uniform \citep{cercignani_book}, thermometers located at different distances from the rotation axis measure a different temperature.

This work completes the axiomatisation of the relativistic thermodynamics proposed by \cite{Noto_full,Israel2009_book}, solving the major open controversies. The formalism was already fully self-consistent (encompassing the first and the second law) and represents the theoretical ground upon which the modern formulations of relativistic hydrodynamics are constructed. Our consistent implementation of the zeroth law provides a direct contact with statistical mechanics and with the standard formulation of thermodynamics.

The take-home message is that heat and friction are the components of a four-vector, representing respectively the exchange of energy and momentum between two thermodynamic bodies. This unification solves the controversy about the relativistic transformation of the heat and, at the same time, changes our understanding of the zeroth law. In fact, since relativity treats energy and momentum symmetrically, so the relativistic thermodynamics will do, adding to the condition of equilibrium with respect to energy exchanges other 3 conditions associated with the exchanges of momentum. This produces in equilibrium 4 constraints, instead of 1, proving that in relativity the zeroth law is still valid and still establishes an equivalence relation. However, the equivalence classes now need to be parametrized using 4 independent parameters (temperature and velocity of the center of mass) instead of just 1.

\section*{Acknowledgements}

 The author thanks M. Antonelli for stimulating discussions and B. Haskell for reading the manuscript and providing critical comments. I acknowledge support from the Polish National Science Centre grant OPUS 2019/33/B/ST9/00942. Partial support comes from PHAROS, COST Action CA16214.

\appendix

\section{Two-level thermometer}\label{2LT}

We consider an ideal non-relativistic Fermi gas enclosed in a box with periodic boundary conditions. Let $C_{\vect{p},\sigma}$ be the annihilation operator of a particle occupying the single-particle state of (discrete) momentum $\vect{p}$ and spin $\sigma=\pm 1/2$. They satisfy the standard anticommutation relations
\begin{equation}
\acomm{ C_{\vect{p},\sigma}}{  C_{\vect{q},\gamma}^{\textcolor{white}{\dagger}} } = \acomm{ C^\dagger_{\vect{p},\sigma}}{  C^\dagger_{\vect{q},\gamma} } = 0
\end{equation}
and
\begin{equation}
\acomm{ C^{\textcolor{white}{\dagger}}_{\vect{p},\sigma}}{  C^\dagger_{\vect{q},\gamma} }  = \delta_{\vect{p},\vect{q}} \delta_{\sigma,\gamma}.
\end{equation}
The Hamiltonian of the gas is
\begin{equation}
H_g = \sum_{\vect{p},\sigma} \epsilon_{\vect{p}} \, C^\dagger_{\vect{p},\sigma}C_{\vect{p},\sigma},
\end{equation}
where $\epsilon_{\vect{p}}$ is the single-particle energy of a fermion of momentum $\vect{p}$. The operator number of particles is
\begin{equation}
N = \sum_{\vect{p},\sigma} C^\dagger_{\vect{p},\sigma} C_{\vect{p},\sigma}.
\end{equation}
We imagine the thermometer to be a two-level system\footnote{We consider a single two-level system for simplicity, but the discussion can be easily generalized to a large number of two-level systems in the thermodynamic limit, producing a rigorously macroscopic object.} with energy separation $\varepsilon>0$ and ground-state energy conventionally set to zero. Therefore the Hamiltonian of the isolated thermometer has the form, in an appropriate basis,
\begin{equation}
H_\tau = \varepsilon \, \sigma_+ \sigma_- = \dfrac{\varepsilon}{2}(1+\sigma_z),
\end{equation}
where $\sigma_\pm$ are constructed as
\begin{equation}
  \sigma_- = \dfrac{1}{2} (\sigma_x -i\sigma_y)  \spc  \sigma_+ =\dfrac{1}{2} (\sigma_x +i\sigma_y) = \sigma_-^\dagger,
\end{equation}
the operators $\sigma_x$, $\sigma_y$, $\sigma_z$ being the Pauli matrices acting on the two-dimensional state space of the thermometer. The total Hilbert space is assumed to be the tensor product between the space of the thermometer and the Fock space of the fermions and we have 
\begin{equation}
\comm{\sigma_j}{C_{\vect{p},\sigma}^{\textcolor{white}{\dagger}}}=0.
\end{equation}
When the thermometer is put into contact with the gas, we assume the total Hamiltonian to be
\begin{equation}
H = H_g + H_\tau + V_{g\tau},
\end{equation} 
where the interaction term $V_{g\tau}$ has the form
\begin{equation}
V_{g\tau}= \sum_{\vect{p},\sigma} g_{\vect{p}} \bigg[ \sigma_+ C_{\vect{p},\sigma} + C^\dagger_{\vect{p},\sigma} \sigma_- \bigg],
\end{equation}
where $g_\vect{p}$ are real coefficients (reassuring Hermitianity). This potential models a process in which the thermometer makes level transitions absorbing and destroying (or creating and emitting) a fermion, breaking the particle number conservation. However it is easy to check that the operator
\begin{equation}
Q =  N + \sigma_+ \sigma_- 
\end{equation} 
commutes with $H$ and therefore describes a constant of motion. In the limit of small $V_{g\tau}$ (weak interaction) the quantity $Z$ introduced in equation \eqref{iorestoferma} is represented by the operator
\begin{equation}
Z= H-\varepsilon \, Q,
\end{equation}
and is, thus, conserved. 

When the system gas+thermometer is in thermal equilibrium, its density matrix can be assumed to be the Generalised Gibbs ensemble \citep{Perarnau_Llobet_2016}
\begin{equation}
\rho \propto \exp(-\beta H +\lambda Q).
\end{equation}
In the limit $V_{g\tau} \rightarrow 0$ it splits into the tensor product
\begin{equation}
\rho = \rho_g \otimes \rho_\tau,
\end{equation}
with
\begin{equation}
\begin{split}
& \rho_g \propto \exp(-\beta H_g +\lambda N) \\
&  \rho_\tau \propto \exp(-\beta H_\tau +\lambda \, \sigma_+ \sigma_-) .\\
\end{split}
\end{equation}
The state $\rho_g$ is the thermal state of the Fermi gas with temperature $T_g$ and chemical potential $\mu$ given by
\begin{equation}\label{asdfgh}
T_g = \beta^{-1}  \spc  \mu = \lambda \beta^{-1}.
\end{equation}
On the other hand, $\rho_\tau$ is the thermal state of the thermometer with temperature
\begin{equation}\label{zxcvbn}
T_\tau = \bigg( \beta - \dfrac{\lambda}{\varepsilon} \bigg)^{-1}.
\end{equation}
Combining \eqref{asdfgh} with \eqref{zxcvbn} we find
\begin{equation}\label{fert}
T_\tau = \dfrac{\varepsilon}{\varepsilon -\mu} T_g,
\end{equation}
which is in agreement with equation \eqref{stranissimo}. Note that, since the gas is a Fermi gas, $\mu$ can in principle be positive. So $T_\tau$ can be infinite (for $\mu = \varepsilon$) or negative (for $\mu > \varepsilon$). In the diluted limit (for $\mu \rightarrow -\infty$) the temperature of the thermometer goes to zero, because the number of fermions is small compared to the number of available single particle states, so the probability to have a particle with exactly the energy $\varepsilon$ to be absorbed is small.

We remark that having $T_\tau \neq T_g$ is possible only as a result of the existence of an other constant of motion apart from the energy, as we clarified in section \ref{neeeeeed}. In fact, if we break the conservation of $Q$, then $\lambda =0$ (so the chemical potential $\mu$ is in turn zero) and \eqref{fert} becomes $T_\tau = T_g$.

\section{A problem of ergodicity}\label{Iosonoegodico}

We have proven that a thermometer can maintain, in equilibrium, a relative motion with the body only if we impose some limitations on the possible exchanges of four-momentum allowed by the interaction, preventing $\delta p^\nu_\tau$ from having arbitrary direction. It is clear that, for macroscopic (i.e. comprised of large numbers of particles) systems with realistic interactions, the existence of an \textit{exact} constant of motion of this kind is nearly impossible to occur. We propose the following thought experiment to explain the basic mechanisms which prevent it.

\subsection{Measuring Planck's and Ott's temperature: an attempt}\label{MPOce}

Consider a real scalar boson $\phi$ of mass $m$, which self-interacts with a short-range (compared with the average distance among the bosons) interaction. Since this particle coincides with its own antiparticle we can assume reactions of the type
\begin{equation}
\phi + \phi \ce{ <=> } \phi + \phi + \phi
\end{equation}
to be allowed, implying that the particle number is not conserved (the chemical potential of the boson vanishes). There are, therefore, no conserved charges and the thermal state of this scalar boson can be characterised by an equation of state of the type \eqref{Quellabuona}. We will consider a gas of bosons $\phi$ to be the body of which we measure the temperature. 

We construct the thermometer as a collection of comoving, with respect to each other, distinguishable two-level systems, where the ground state is denoted by $\ket{0}$ and the excited state is denoted by $\ket{1}$.  We impose the mass separation between the two levels to be equal to the mass of the boson:
\begin{equation}\label{deltaM}
M_{\ket{1}}-M_{\ket{0}} =m.
\end{equation}
The interaction between the thermometer and the gas is assumed to happen only through the absorption or the emission of a boson $\phi$ by a two-level system, namely through a reaction
\begin{equation}\label{termiamo}
B_{\ket{0}} + \phi \ce{ <=> } B_{\ket{1}},
\end{equation}  
where $B_{\ket{0}}$ represents a two-level system in the ground state and $B_{\ket{1}}$ represents a two-level system in the excited state. Equation \eqref{deltaM} implies that the reaction \eqref{termiamo} can happen only in the case in which the boson $\phi$ is at rest with respect to the thermometer, 
to conserve the total four-momentum. Therefore the reaction alters the mass of the thermometer but not its state of motion. We have constructed a thermometer whose four-velocity $u_\tau^\nu$ is conserved by the interaction with the body. 

According to the statistical arguments we exposed in subsection \ref{qpwodioemfowemfmo}, this thermometer will measure, in equilibrium, Planck's temperature. This can be also shown directly from kinetic theory. In fact, if we work in the reference frame of the thermometer, it is easy to show that the number $N_{\ket{1}}$ of two-level systems in the excited state evolves according to the equation \citep{dirac1981principles}
\begin{equation}\label{gkgkg}
\dfrac{dN_{\ket{1}}}{dt} = \mathcal{R} \bigg[ N_{\ket{0}} n_0 - N_{\ket{1}} (1+n_0)  \bigg],
\end{equation} 
where $\mathcal{R}$ is the spontaneous decay rate of the state $\ket{1}$, $N_{\ket{0}}$ is the number of two-level systems in the ground state and $n_{\vect{q}}$ is the average number of bosons $\phi$ with spatial momentum $\vect{q}$. Once thermal equilibrium is reached, $N_{\ket{1}}$ becomes constant and from \eqref{gkgkg} we obtain
\begin{equation}\label{hjk}
\dfrac{N_{\ket{1}}}{N_{\ket{0}}} = \dfrac{n_0}{1+n_0}.
\end{equation}
Recalling that the interaction is short range, we can approximate the gas as ideal. The equilibrium occupation number $n_{\vect{q}}$ of the ideal boson gas in a thermal state is 
\begin{equation}
n_\vect{q} = \dfrac{1}{e^{-\beta_\nu q^\nu}-1},
\end{equation} 
thus equation \eqref{hjk} becomes
\begin{equation}
\dfrac{N_{\ket{1}}}{N_{\ket{0}}} = e^{-\beta^0 m},
\end{equation} 
which describes the macrostate of the thermometer having temperature
\begin{equation}
T_\tau = \dfrac{1}{\beta^0} .
\end{equation}
Recalling equation \eqref{vnidfovkm}, we see that the above equilibrium condition reduces to Planck's prescription. Thus, this thought experiment might seem, at the first sight, to support equation \eqref{Planck} as the transformation law of the temperature under Lorentz boosts. A supporter of Ott's view, however, performing the same experiment might consider the collection of two-level systems to be body of which they are measuring the temperature and the boson gas as the thermometer. They will therefore exchange $T_\tau$ with $T_{RF}$, obtaining a measurement of the temperature of the two-level system in agreement with Ott's law, equation \eqref{Ott}.

\subsection{Realistic interactions}\label{mareale}

The situation we have  proposed represents an idealised system in which the thermometer can exchange energy but not momentum (in its reference frame). Unfortunately such a system cannot be constructed in realistic conditions and represents only a non-physical limit. In fact the spontaneous decay rate $\mathcal{R}$ appearing in equation \eqref{gkgkg} is given by the \textit{Fermi Golden Rule} and must be equal to \citep{Peskin_book}
\begin{equation}
\mathcal{R} = \dfrac{|\vect{q}_f|}{8\pi {M_{\ket{1}}}^2} |\bra{0,\phi}V_I\ket{1}|^2,
\end{equation}
where  $|\bra{0,\phi}V\ket{1}|^2$ is the matrix element of the interaction potential responsible for the transition and $\vect{q}_f$ is the momentum of the emitted particle $\phi$ in the original frame of the two-level system. So if we set $\vect{q}_f=0$, which is the necessary condition to guarantee the conservation of $u_\tau^\nu$, we will find that the reaction cannot occur and the thermometer does not interact with the body at all. The only way to make the transfusion of energy possible is to set $|\vect{q}_f|\neq 0$, breaking the conservation of $u_\tau^\nu$. The typical acceleration that the thermometer will, then, experience is 
\begin{equation}
a_c \sim \mathcal{R}\dfrac{|\vect{q}_f|}{M_{\ket{0}}},
\end{equation}
which is a second order in $|\vect{q}_f|/M_{\ket{0}}$. Therefore the system presented in subsection \ref{MPOce} can exist only as a limit in which the time required for the thermometer to reach comotion with the body is much longer than the time of the experiment. 

There is, however, an other complication. Interactions coming from an underlying field theory in general admit the possibility of having higher order processes, of the kind
\begin{equation}\label{termiamo33}
B_{\ket{0}} + \phi +\phi \ce{ <=> } B_{\ket{1}}+\phi,
\end{equation} 
where a spectator $\phi$ takes part of the four-momentum. In this situation the two-level system does not need to be at rest at the end of the process, breaking again the conservation of $u_\tau^\nu$. This process is not suppressed by the phase space and therefore becomes dominant, leading anyway to the inevitable acceleration of the thermometer.

\section{Ambiguity in the definition of the heat}\label{Calormio}

Let us start from the general differential
\begin{equation}\label{dividotutto}
\delta p^\nu = u^\nu \delta M + M \delta u^\nu.
\end{equation}
This represents an orthogonal decomposition of the variation of the four-momentum, because
\begin{equation}
u_{\nu} \delta u^\nu =0, 
\end{equation}
which is ensured by the conservation of the normalization condition $u_\nu u^\nu =-1$. Let us assume an equation of state \eqref{Quellabuona}, which implies that we do not have chemical-type forms of work $\mu^A \delta Q_A$ (where $\mu^A$ is the chemical potential of the charge $Q_A$), but the only possible way to exert work on the body is by impressing an acceleration through the action of a force. This suggests a relativistic first-law
\begin{equation}\label{dnvoekmcow}
\delta p^\nu = \delta \mathcal{Q}^\nu + \delta \mathcal{W}^\nu
\end{equation}
with
\begin{equation}\label{dQdW}
\delta \mathcal{Q}^\nu = u^\nu \delta M  \spc \delta \mathcal{W}^\nu = M \delta u^\nu.
\end{equation}
The four-vector $\delta \mathcal{W}^\nu$ can be interpreted as the four-dimensional generalization of the work element. In fact, in a given reference frame, 
\begin{equation}\label{W1O}
\delta \mathcal{W}^0 =(u^0)^3 M v_j \delta v^j,
\end{equation}
which is the formula for the relativistic work of an external force acting on a moving particle. From equation \eqref{dQdW} one finds that in a given frame
\begin{equation}
\delta \mathcal{Q}^0 =u^0 \delta M= u^0 T_{RF} \delta S,
\end{equation} 
where we have employed the definition \eqref{TRFFF}. This formula is the relativistic transformation of the heat proposed by Ott \citep{Pinto2017}.

There is, however, an other possibility. Let us rewrite equation \eqref{Massa} in a chosen reference frame in the form
\begin{equation}
E = \sqrt{M^2 + p_j p^j}.
\end{equation}
Its differential reads
\begin{equation}
\delta E = \dfrac{\delta M}{u^0} + v^j \delta p_j.
\end{equation}
The second term reminds the standard formula of the variation of the kinetic energy provided in Hamiltonian mechanics, which then suggests the decomposition
\begin{equation}
\delta E = \delta \mathcal{Q} +\delta \mathcal{W},
\end{equation}
with
\begin{equation}\label{W2P}
\delta \mathcal{W} = v^j \delta p_j
\end{equation}
and
\begin{equation}
\delta \mathcal{Q} = \dfrac{\delta M}{u^0} = \dfrac{T_{RF} \delta S}{u^0}.
\end{equation}
This is the relativistic transformation for the heat proposed by Planck \citep{Pinto2017}. We see that the source of ambiguity is in the definition of the work as the variation of the kinetic energy of the system in situations in which the mass is changing with time (in the case in which $\delta M =0$ the formulas of the work \eqref{W1O} and \eqref{W2P} coincide, as it happens in point-particle mechanics). 

As we clarify in the main text, none of these two subdivisions reflects the standard physical interpretation of the splitting of the energy variation into heat and work. The subdivision needs to be performed considering how the environment is acting on the body and cannot be uniquely determined in terms of internal properties of the body itself.

\bibliographystyle{mnras}
\bibliography{Biblio}

\begin{thebibliography}{}
\makeatletter
\relax
\def\mn@urlcharsother{\let\do\@makeother \do\$\do\&\do\#\do\^\do\_\do\%\do\~}
\def\mn@doi{\begingroup\mn@urlcharsother \@ifnextchar [ {\mn@doi@}
  {\mn@doi@[]}}
\def\mn@doi@[#1]#2{\def\@tempa{#1}\ifx\@tempa\@empty \href
  {http://dx.doi.org/#2} {doi:#2}\else \href {http://dx.doi.org/#2} {#1}\fi
  \endgroup}
\def\mn@eprint#1#2{\mn@eprint@#1:#2::\@nil}
\def\mn@eprint@arXiv#1{\href {http://arxiv.org/abs/#1} {{\tt arXiv:#1}}}
\def\mn@eprint@dblp#1{\href {http://dblp.uni-trier.de/rec/bibtex/#1.xml}
  {dblp:#1}}
\def\mn@eprint@#1:#2:#3:#4\@nil{\def\@tempa {#1}\def\@tempb {#2}\def\@tempc
  {#3}\ifx \@tempc \@empty \let \@tempc \@tempb \let \@tempb \@tempa \fi \ifx
  \@tempb \@empty \def\@tempb {arXiv}\fi \@ifundefined
  {mn@eprint@\@tempb}{\@tempb:\@tempc}{\expandafter \expandafter \csname
  mn@eprint@\@tempb\endcsname \expandafter{\@tempc}}}

\bibitem[\protect\citeauthoryear{{Andersson} \& {Comer}}{{Andersson} \&
  {Comer}}{2007}]{andersson2007review}
{Andersson} N.,  {Comer} G.~L.,  2007, \mn@doi [Living Reviews in Relativity]
  {10.12942/lrr-2007-1}, \href
  {http://adsabs.harvard.edu/abs/2007LRR....10....1A} {10, 1}

\bibitem[\protect\citeauthoryear{{Becattini}}{{Becattini}}{2016}]{Becattini2016}
{Becattini} F.,  2016, \mn@doi [Acta Physica Polonica B]
  {10.5506/APhysPolB.47.1819}, \href
  {https://ui.adsabs.harvard.edu/abs/2016AcPPB..47.1819B} {47, 1819}

\bibitem[\protect\citeauthoryear{{B{\'\i}r{\'o}} \& {V{\'a}n}}{{B{\'\i}r{\'o}}
  \& {V{\'a}n}}{2010}]{Biro2010}
{B{\'\i}r{\'o}} T.~S.,  {V{\'a}n} P.,  2010, \mn@doi [EPL (Europhysics
  Letters)] {10.1209/0295-5075/89/30001}, \href
  {https://ui.adsabs.harvard.edu/abs/2010EL.....8930001B} {89, 30001}

\bibitem[\protect\citeauthoryear{Callen}{Callen}{1985}]{Callen_book}
Callen H.~B.,  1985, {Thermodynamics and an introduction to thermostatistics;
  2nd ed.}.
Wiley, New York, NY, \url {https://cds.cern.ch/record/450289}

\bibitem[\protect\citeauthoryear{Carath\'{e}odory}{Carath\'{e}odory}{1909}]{Caratheodory1909}
Carath\'{e}odory C.,  1909, Mathematische Annalen, 67, 355

\bibitem[\protect\citeauthoryear{Carter}{Carter}{1989}]{noto_rel}
Carter B.,  1989, Covariant theory of conductivity in ideal fluid or solid
  media.
 Vol. 1385, \mn@doi{10.1007/BFb0084028, }

\bibitem[\protect\citeauthoryear{{Carter}}{{Carter}}{1991}]{carter1991}
{Carter} B.,  1991, \mn@doi [Proceedings of the Royal Society of London Series
  A] {10.1098/rspa.1991.0034}, \href
  {https://ui.adsabs.harvard.edu/abs/1991RSPSA.433...45C} {433, 45}

\bibitem[\protect\citeauthoryear{Carter \& Khalatnikov}{Carter \&
  Khalatnikov}{1992}]{CarterKhal_equivalence}
Carter B.,  Khalatnikov I.~M.,  1992, \mn@doi [Phys. Rev. D]
  {10.1103/PhysRevD.45.4536}, 45, 4536

\bibitem[\protect\citeauthoryear{{Cercignani} \& {Kremer}}{{Cercignani} \&
  {Kremer}}{2002}]{cercignani_book}
{Cercignani} C.,  {Kremer} G.~M.,  2002, {The relativistic Boltzmann equation:
  theory and applications}

\bibitem[\protect\citeauthoryear{De~Groot}{De~Groot}{1980}]{degroot_book}
De~Groot S.,  1980, {Relativistic Kinetic Theory. Principles and Applications}

\bibitem[\protect\citeauthoryear{{Dirac}}{{Dirac}}{1947}]{dirac1981principles}
{Dirac} P.~A.~M.,  1947, {The principles of quantum mechanics}

\bibitem[\protect\citeauthoryear{{Far{\'\i}as}, {Pinto}  \&
  {Moya}}{{Far{\'\i}as} et~al.}{2017}]{Pinto2017}
{Far{\'\i}as} C.,  {Pinto} V.~A.,   {Moya} P.~S.,  2017, \mn@doi [Scientific
  Reports] {10.1038/s41598-017-17526-4}, \href
  {https://ui.adsabs.harvard.edu/abs/2017NatSR...717657F} {7, 17657}

\bibitem[\protect\citeauthoryear{Florkowski}{Florkowski}{2010}]{Florkowski:2010zz}
Florkowski W.,  2010, {Phenomenology of Ultra-Relativistic Heavy-Ion
  Collisions}

\bibitem[\protect\citeauthoryear{Gavassino \& Antonelli}{Gavassino \&
  Antonelli}{2019}]{Termo}
Gavassino L.,  Antonelli M.,  2019, \mn@doi [Classical and Quantum Gravity]
  {10.1088/1361-6382/ab5f23}, 37, 025014

\bibitem[\protect\citeauthoryear{{Gavassino}, {Antonelli}  \&
  {Haskell}}{{Gavassino} et~al.}{2020}]{Gavassino2020Bulk}
{Gavassino} L.,  {Antonelli} M.,   {Haskell} B.,  2020, arXiv e-prints, \href
  {https://ui.adsabs.harvard.edu/abs/2020arXiv200304609G} {p. arXiv:2003.04609}

\bibitem[\protect\citeauthoryear{Guggenheim}{Guggenheim}{1985}]{Guggenheim1986}
Guggenheim E.~A.,  1985, {Thermodynamics - An advanced treatment for chemists
  and physicists (7th edition)}

\bibitem[\protect\citeauthoryear{Huang}{Huang}{1987}]{huang_book}
Huang K.,  1987, Statistical Mechanics, 2 edn.
John Wiley \& Sons

\bibitem[\protect\citeauthoryear{Israel}{Israel}{1981}]{Israel_1981_review}
Israel W.,  1981, \mn@doi [Physica A: Statistical Mechanics and its
  Applications] {https://doi.org/10.1016/0378-4371(81)90220-X}, 106, 204

\bibitem[\protect\citeauthoryear{Israel}{Israel}{1989}]{Noto_full}
Israel W.,  1989, in Anile A.~M.,  Choquet-Bruhat Y.,  eds, Relativistic Fluid
  Dynamics. Springer Berlin Heidelberg, Berlin, Heidelberg, pp 152--210

\bibitem[\protect\citeauthoryear{Israel}{Israel}{2009}]{Israel2009_book}
Israel W.,  2009, Relativistic Thermodynamics.
Birkh{\"a}user Basel, Basel, pp 101--113, \mn@doi{10.1007/978-3-7643-8878-2_8},
  \url {https://doi.org/10.1007/978-3-7643-8878-2_8}

\bibitem[\protect\citeauthoryear{Israel \& Stewart}{Israel \&
  Stewart}{1979}]{Israel_Stewart_1979}
Israel W.,  Stewart J.,  1979, \mn@doi [Annals of Physics]
  {https://doi.org/10.1016/0003-4916(79)90130-1}, 118, 341

\bibitem[\protect\citeauthoryear{Jaynes}{Jaynes}{1957a}]{Jaynes1}
Jaynes E.~T.,  1957a, \mn@doi [Phys. Rev.] {10.1103/PhysRev.106.620}, 106, 620

\bibitem[\protect\citeauthoryear{Jaynes}{Jaynes}{1957b}]{Jaynes2}
Jaynes E.~T.,  1957b, \mn@doi [Phys. Rev.] {10.1103/PhysRev.108.171}, 108, 171

\bibitem[\protect\citeauthoryear{Khinchin}{Khinchin}{1949}]{khinchin_book}
Khinchin A.,  1949, Mathematical Foundations of Statistical Mechanics.
Dover Books on Mathematics, Dover Publications, \url
  {https://books.google.pl/books?id=i6HbVpZclcoC}

\bibitem[\protect\citeauthoryear{Landau \& Lifshitz}{Landau \&
  Lifshitz}{2013}]{landau5}
Landau L.,  Lifshitz E.,  2013, Statistical Physics.
No.~v. 5, Elsevier Science, \url
  {https://books.google.pl/books?id=VzgJN-XPTRsC}

\bibitem[\protect\citeauthoryear{{Landsberg}}{{Landsberg}}{1967}]{Landsberg1967_cool}
{Landsberg} P.~T.,  1967, \mn@doi [\nat] {10.1038/214903a0}, \href
  {https://ui.adsabs.harvard.edu/abs/1967Natur.214..903L} {214, 903}

\bibitem[\protect\citeauthoryear{Landsberg \& Matsas}{Landsberg \&
  Matsas}{1996}]{Landsberg_1996}
Landsberg P.~T.,  Matsas G.~E.,  1996, \mn@doi [Physics Letters A]
  {10.1016/s0375-9601(96)00791-8}, 223, 401

\bibitem[\protect\citeauthoryear{Landsberg \& Matsas}{Landsberg \&
  Matsas}{2004}]{Landsberg2004}
Landsberg P.,  Matsas G.,  2004, \mn@doi [Physica A: Statistical Mechanics and
  its Applications] {https://doi.org/10.1016/j.physa.2004.03.081}, 340, 92

\bibitem[\protect\citeauthoryear{{Lopez-Monsalvo} \&
  {Andersson}}{{Lopez-Monsalvo} \& {Andersson}}{2011}]{lopez2011}
{Lopez-Monsalvo} C.~S.,  {Andersson} N.,  2011, \mn@doi [Proceedings of the
  Royal Society of London Series A] {10.1098/rspa.2010.0308}, \href
  {https://ui.adsabs.harvard.edu/\#abs/2011RSPSA.467..738L} {467, 738}

\bibitem[\protect\citeauthoryear{{Mare{\v{s}}}, Hub\'{i}k  \&
  Spicka}{{Mare{\v{s}}} et~al.}{2017}]{Mares2017}
{Mare{\v{s}}} J.,  Hub\'{i}k P.,   Spicka V.,  2017, \mn@doi [Fortschritte der
  Physik] {10.1002/prop.201700018}, 65, 1700018

\bibitem[\protect\citeauthoryear{Oppenheim}{Oppenheim}{2003a}]{Oppenheim2003}
Oppenheim J.,  2003a, \mn@doi [Phys. Rev. E] {10.1103/PhysRevE.68.016108}, 68,
  016108

\bibitem[\protect\citeauthoryear{Oppenheim}{Oppenheim}{2003b}]{Oppenheim2003B}
Oppenheim J.,  2003b, \mn@doi [Phys. Rev. E] {10.1103/PhysRevE.68.016108}, 68,
  016108

\bibitem[\protect\citeauthoryear{{Ott}}{{Ott}}{1963}]{Ott1963}
{Ott} H.,  1963, \mn@doi [Zeitschrift fur Physik] {10.1007/BF01375397}, \href
  {https://ui.adsabs.harvard.edu/abs/1963ZPhy..175...70O} {175, 70}

\bibitem[\protect\citeauthoryear{Parisi}{Parisi}{1988}]{Parisi_book_1988}
Parisi G.,  1988, {STATISTICAL FIELD THEORY}

\bibitem[\protect\citeauthoryear{Parvan}{Parvan}{2019}]{PARVAN2019}
Parvan A.,  2019, \mn@doi [Annals of Physics]
  {https://doi.org/10.1016/j.aop.2019.01.003}, 401, 130

\bibitem[\protect\citeauthoryear{Perarnau-Llobet, Riera, Gallego, Wilming  \&
  Eisert}{Perarnau-Llobet et~al.}{2016}]{Perarnau_Llobet_2016}
Perarnau-Llobet M.,  Riera A.,  Gallego R.,  Wilming H.,   Eisert J.,  2016,
  \mn@doi [New Journal of Physics] {10.1088/1367-2630/aa4fa6}, 18, 123035

\bibitem[\protect\citeauthoryear{Peskin \& Schroeder}{Peskin \&
  Schroeder}{1995}]{Peskin_book}
Peskin M.~E.,  Schroeder D.~V.,  1995, {An Introduction to quantum field
  theory}.
Addison-Wesley, Reading, USA, \url
  {http://www.slac.stanford.edu/~mpeskin/QFT.html}

\bibitem[\protect\citeauthoryear{Petz}{Petz}{2001}]{VonNeumannEntropy2001}
Petz D.,  2001, Entropy, von Neumann and the von Neumann Entropy.
Springer Netherlands, Dordrecht, pp 83--96,
  \mn@doi{10.1007/978-94-017-2012-0_7}, \url
  {https://doi.org/10.1007/978-94-017-2012-0_7}

\bibitem[\protect\citeauthoryear{Planck}{Planck}{1908}]{Planck_1908}
Planck M.,  1908, \mn@doi [Annalen der Physik] {10.1002/andp.19083310602}, 331,
  1

\bibitem[\protect\citeauthoryear{{Rezzolla} \& {Zanotti}}{{Rezzolla} \&
  {Zanotti}}{2013}]{rezzolla_book}
{Rezzolla} L.,  {Zanotti} O.,  2013, {Relativistic Hydrodynamics}

\bibitem[\protect\citeauthoryear{{Sewell}}{{Sewell}}{2008}]{Sewell2008}
{Sewell} G.~L.,  2008, \mn@doi [Journal of Physics A Mathematical General]
  {10.1088/1751-8113/41/38/382003}, \href
  {https://ui.adsabs.harvard.edu/abs/2008JPhA...41L2003S} {41, 382003}

\bibitem[\protect\citeauthoryear{{Tolman}}{{Tolman}}{1933}]{Tolman1933}
{Tolman} R.~C.,  1933, \mn@doi [Science] {10.1126/science.77.1995.291}, \href
  {https://ui.adsabs.harvard.edu/abs/1933Sci....77..291T} {77, 291}

\bibitem[\protect\citeauthoryear{{Weinberg}}{{Weinberg}}{1972}]{Weinberg_book_1972}
{Weinberg} S.,  1972, {Gravitation and Cosmology: Principles and Applications
  of the General Theory of Relativity}

\bibitem[\protect\citeauthoryear{van Kampen}{van Kampen}{1968}]{vanKampen1968}
van Kampen N.~G.,  1968, \mn@doi [Phys. Rev.] {10.1103/PhysRev.173.295}, 173,
  295

\bibitem[\protect\citeauthoryear{van Weert}{van Weert}{1982}]{VanWeert1982}
van Weert C.,  1982, \mn@doi [Annals of Physics]
  {https://doi.org/10.1016/0003-4916(82)90338-4}, 140, 133

\makeatother
\end{thebibliography}

\end{document}